\documentclass[aps,prl,twocolumn,floats]{revtex4}
\usepackage{graphicx}
\begin{document}
\title{Wave nucleation rate in excitable systems in the low noise limit}
\author{Herv\'e Henry and Herbert Levine}
\date{\today}
\affiliation{Center for Theoretical Biological Physics\\ University
of California San Diego}
\begin{abstract}
Motivated by recent experiments on intracellular calcium dynamics, we
study the general issue of fluctuation-induced nucleation of waves in
excitable media. We utilize a stochastic Fitzhugh-Nagumo model for
this study, a spatially-extended non-potential pair of equations
driven by thermal (i.e. white) noise. The nucleation rate is
determined by finding the most probable escape path via minimization
of an action related to the deviation of the fields from their
deterministic trajectories. Our results pave the way both for studies
of more realistic models of calcium dynamics as well as of nucleation
phenomena in other non-equilibrium pattern-forming processes.
\end{abstract}

\maketitle

One very important class of non-equilibrium spatially-extended
systems is that of excitable media. In these, a quiescent state is
linearly stable but nonlinear waves can nonetheless propagate without
decaying. These waves can be generated by above-threshold local
perturbations and they also can become self-sustaining in the form of
rotating spirals\cite{Win72}.
Examples of excitable media include many biological
systems such as the cAMP waves seen in Dictyostelium amoebae
aggregation\cite{Loo}, electrical waves in cardiac and neural
tissue\cite{WTBD} and, primary for our focus here, intracellular calcium
waves\cite{Berridge}.

Most excitable systems are sufficiently macroscopic as to render
unimportant the role of thermodynamic fluctuations and allow for a
description in terms of deterministic pde models. For these cases,
noise effects can still be studied via the imposition of external
variation in time and/or space (e.g. by varying illumination in a
light-sensitive BZ reaction \cite{ShowalterPRL,ShowalterChaos});
however, there is no need to include noise in a description of the
``natural" version of these systems.
\begin{figure}
\begin{center}
\includegraphics[width=3.5cm]{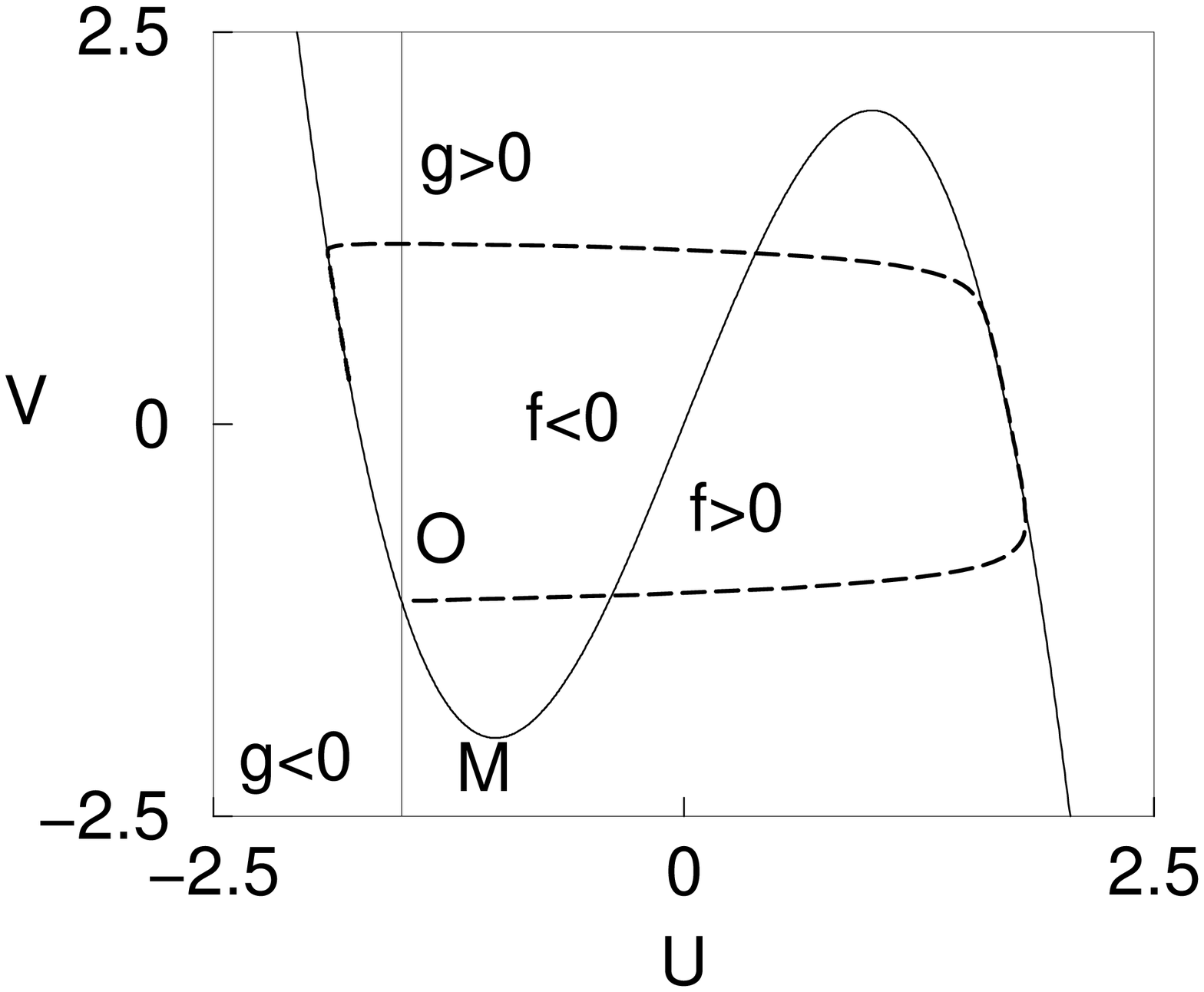}
\includegraphics[width=3.5cm,height=3.cm]{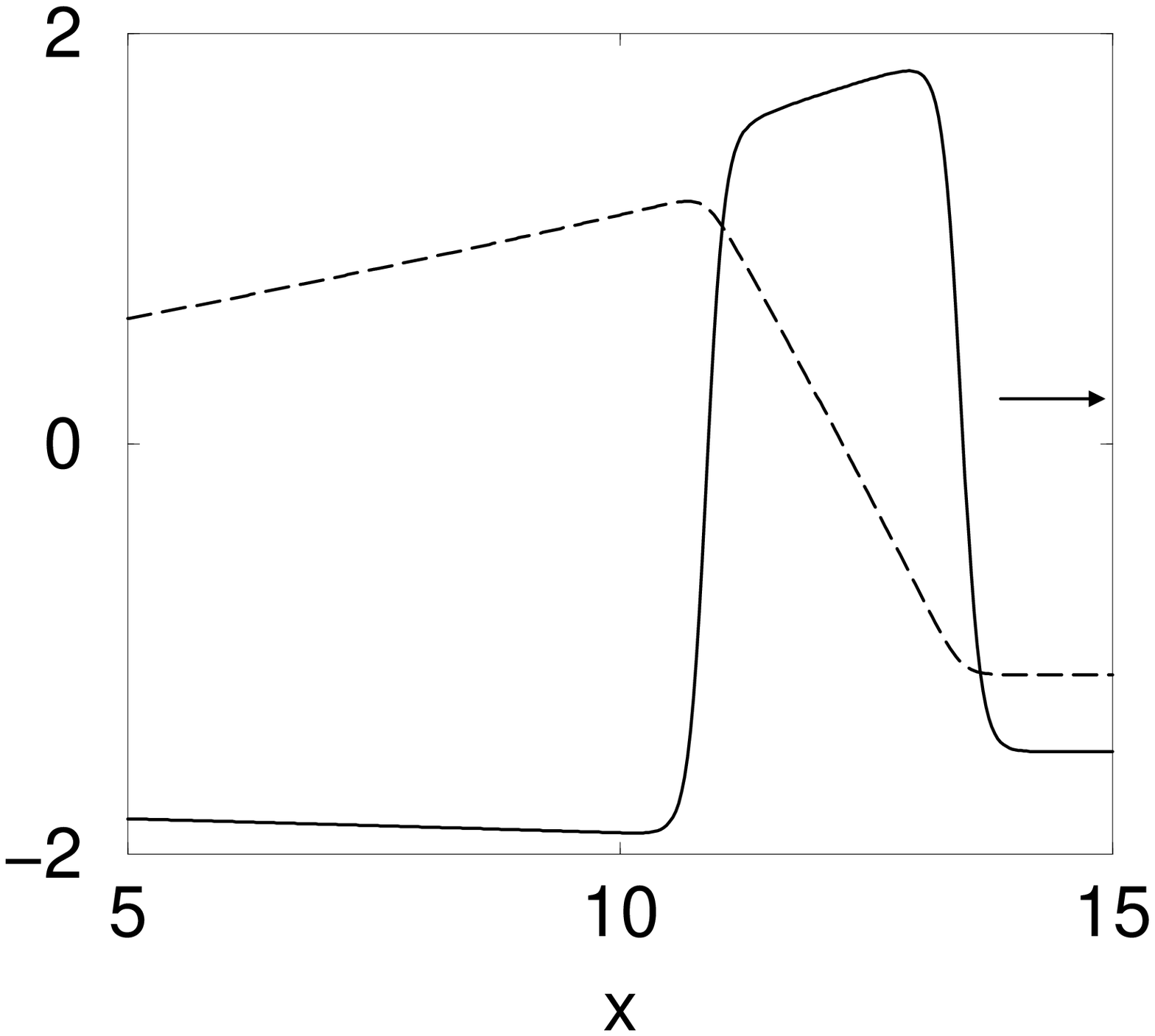}
\caption{\label{FHN} Left: Phase space diagram of the Fitzhugh-Nagumo
system without noise and without the diffusive term. The stable
equilibrium point is $O$. The $f(u,v)=3u-u^3-v$ nullcline is the S
shaped solid line with minimum at $M$ and the $g=u+(1+\gamma )$
nullcline is the thin solid line. A small perturbation of $O$ leads
to a large excursion in phase space. The dashed line shows the
phase-space excursion of a point during the propagation of a single
wave. Parameter values are $\gamma=.5$ and $\epsilon=0.05$. Right:
Typical propagating wave in an excitable medium. Solid line: $u$;
dashed line: $v$. The widths of the wave front and wave back (regions
of fast change in $u$ where $(u,\,v)$ is not on the $f$ nullcline)
are of order $\sqrt{\epsilon}$. The full return to equilibrium  is
not shown.}
\end{center}
\end{figure}
This is manifestly not the case for intracellular calcium dynamics;
since the excitability here arises through the opening and closing of
a small number of ion channels (allowing/preventing calcium efflux
from calcium stores\cite{Shuai}), fluctuations are inherently
non-negligible. Indeed, experiments show direct evidence of noise
effects in the form of abortive waves and spontaneous wave
nucleation\cite{Llano99,Parker}.

In this paper, we study the process of spontaneous wave nucleation
for a 1d generic excitable system modeled by the Fitzhugh-Nagumo equations
\begin{eqnarray}
\partial_t u &=&(3u-u^3-v)/\epsilon+\nabla^2 u+\eta_u\\
\partial_t v &=& u+(1+\gamma) +\eta_v
\end{eqnarray}
\label{FHNequation} Here $\eta_v$ and $\eta_u$ are small independent
white noise terms  modeling fluctuation effects with covariance equal
to:
\begin{equation}
 <\eta_i(x,t)\eta_j(x',t')> = \beta \delta _{ij} \delta(x-x')\delta(t-t')
\end{equation}
At $\beta=0$ and for positive values of $\gamma$, this system is
excitable with a single stable equilibrium point $u_0=-(1+\gamma),\
v_0=3u_0-u_0^3$. As already mentioned, a wave of excitation can
propagate through the system (see fig.\ref{FHN}); a
counter-propagating pair of such waves will be generated if a local
perturbation above a threshold value is applied. For negative values
of $\gamma$, the system becomes oscillatory. This model is not meant
to be a realistic approximation for any specific physical or
biological process; instead we use the model to understand the
generic features of wave nucleation due to noise.

\begin{figure}
\begin{center}
\includegraphics[width=4.5cm]{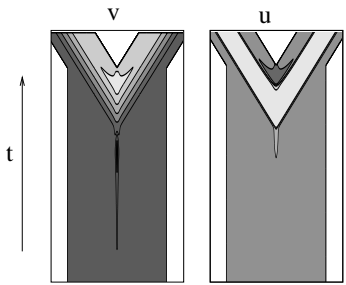}
\caption{\label{DessinMPEP1}  Space time contour plot of the $u$
(right) and $v$(left) field during nucleation with $T$=20; parameter
values are $\epsilon=0.1$ and $\gamma=.2$.  Time increases upward and
the white area corresponds to the region outside of
$[x_{min}(t)=max(x_f(t)-x_{off},0),x_{max}(t)=
max(L+x_f(t)-x_{off},0)],\ [0,T]$. The contours are at   -2, -1, 0,
and  1 for the  $u$ field and at $v_0-0.2$,  $v_0-0.05$, $v_0+0.05$,
$v_0+1$, $v_0+2$ and $v_0+3$  for the $v$ field. The lighter the
surface, the higher the values of the fields are; thus, the light
region of the $u$ contour plot corresponds to the excited region.
After the nucleation event, points near the center reach the maximum
of the excited branch of the $f$ nullcline (see fig. \ref{FHN}),
return to the quiescent state and thereby give birth to the two
wave-backs.}
\end{center}
\end{figure}
\begin{figure}
\begin{small}
\begin{tabular}{cc}
\hspace{1.2cm}(a)&\hspace{1.2cm} (b)\\
\includegraphics[width=3.5cm]{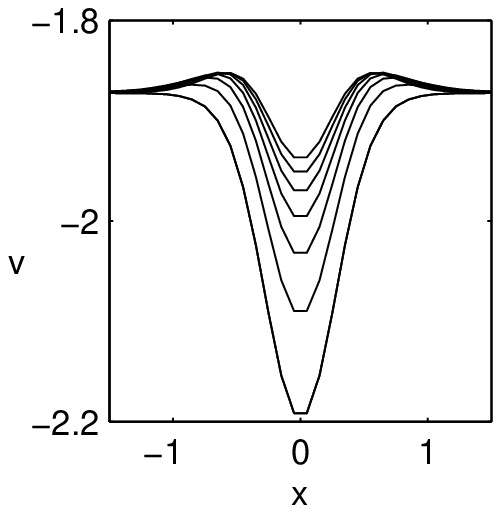}&
\includegraphics[width=3.5cm]{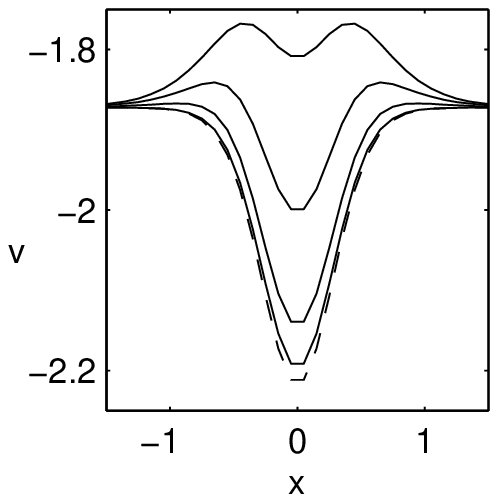}\\
\hspace{1.2cm}(c)&\hspace{1.2cm} (d)\\
\includegraphics[width=3.5cm]{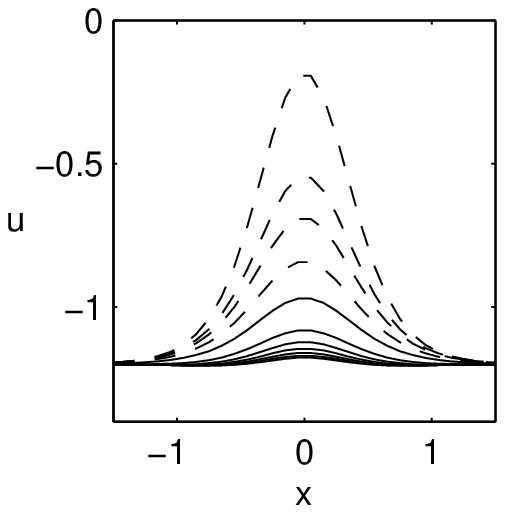}&
\includegraphics[width=3.5cm]{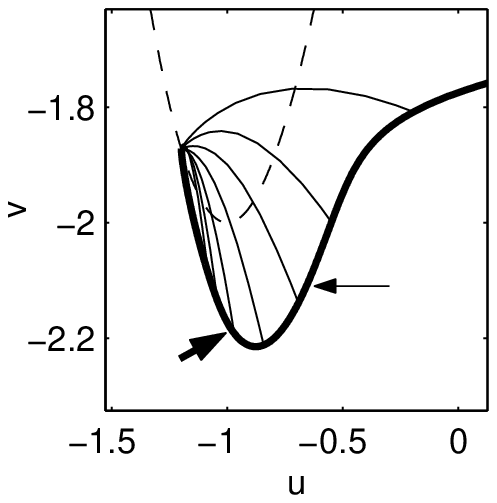}
\end{tabular}
\end{small}
\caption{\label{DessinMPEP2}(a) and (b) $v$ profiles during a typical
nucleation event.  (a)  $v$ profiles at times 25, 25.5, 26, 26.5, 27,
27.5 and 28. During that stage, the minimal value of the $v$ field
decreases. (b) $v$ profiles at times 28(dashed), 28.25, 28.5, 28.75
and 29 (solid lines). During this stage, the minimal value of $v$
increases. (c): u profiles at the same times. The profiles
corresponding to the (a),(b) curves are represented with a solid,
dashed line. (d): The dashed line represents the $f$ nullcline. The
thick solid line is the trajectory in the phase plane
($u(x=0,t),v(x=0,t)$) of the center of nucleation. The point where
maximum value of $\int dx \delta_u^2+\delta_v^2$ is reached is
indicated by a thick arrow, whereas the point where this deviation is
below 1\% of this maximal value is indicated by the thin arrow.
Beyond this point the dynamics are mainly driven by the deterministic
equations. The thin solid lines represent spatial snapshots of the
MPEP ($u(x,t=T_i),\ v(x,t=T_i)$) profile at regularly spaced points
in time ($T_i-T_{i-1}= 0.5$). }
\end{figure}
Specifically, at finite $\beta$ the noise will allow the birth of
pairs of counter-propagating wave at a rate that will depend on the
amplitude of the noise. That rate can be determined in the case of
relatively high noise using  direct numerical simulations. However,
in the case of low noise, such a method becomes computationally
prohibitive. Here we present a computation of the transition rate
using a most probable escape path (MPEP) approach that is based on
solving the Fokker-Planck equation \cite{Krammers1940,Freidlin}. This
method has been successfully applied to dynamical
systems\cite{Graham,MaierStein93} and to a few cases of spatially
extended systems, namely the transition from creep to
fracture\cite{Marder95} and magnetic domain
reversal\cite{Weinan2002}. To derive the equation for the MPEP, we
use the fact that the solution of the Fokker-Planck equation with
given initial and final field configurations for time interval
$[0,T]$ can be written in terms of the path integral
\begin{equation}
P(T) = \int\mathcal{ D}(u,v) \exp\left(-
\frac{1}{\beta}\int\int\,dtdx\ S(x,t)  \right)\label{Pathint}
\\
\end{equation}
with the ``action density" $S$ given by the sum of the squared
deviations of the time derivatives of the fields from their
deterministic values: $S \equiv \delta_u ^2 +\delta _v ^2$
\begin{eqnarray}
\delta_u&=&\partial_t u -(3u-u^3-v)/\epsilon-\Delta u \\
\delta_v&=&\partial_t v -u-(1+\gamma)
\end{eqnarray}
 The functional integral is taken over all paths that begin at
$t=0$ in equilibrium and end with a given final counter-propagating
wave state  $(u_f,v_f)$ at $t=T$. Since we take $\beta$ close to 0,
the r.h.s. of eq. (\ref{Pathint}) is dominated by the path (called
most probable escape path or MPEP) that maximizes the integrand over
all paths; thus, the transition rate is found to be proportional to:
\begin{equation}
\exp(-E/\beta)\label{Pbte_trans}
\end{equation}
where E is the minimum over all paths of $ \int dx\int dt \, S$

Therefore, in order to compute the transition rate between the rest
state and a pair of counter propagating waves, one has only to
compute the minimum of the quantity in eq. (\ref{Pathint}). This
minimum can be expressed using a variational principle as the
solution of a PDE\cite{Marder96}; however, using that approach to
finding the actual MPEP between two different states involves the use
of a shooting method with numerous parameters, which turns out to be
numerically quite difficult. Instead, we used an alternative method
based on discretizing the above path integral on a space-time grid
and directly using a quasi-Newton\cite{Nocedal} method to find the
minimum. One difficulty with this approach is that the Fitzhugh
Nagumo model has a slow recovery  time compared to the timescale
associated with a pulse (width of pulse/speed). This then
necessitates having a very large spatial domain, if one attempts to
completely encompass the region over which the nucleated wave
configuration differs from the quiescent fixed point. To get around
this difficulty, we use a grid moving with the pulse. That is, the
grid moves  in the same direction as the pulse in order to keep the
wave front at a fixed  distance from the boundary. Thus, the
boundaries of the domain used to compute the MPEP path are no longer
$[0,\,L],\, [0,T]$, but $[x_{min}(t)=0+\mbox{max}(x_f(t)-x_{off},0),\
x_{max}(t)= L+\mbox{max}(x_f(t)-x_{off},0)],\, [0,T]$, where $x_f(t)$
is the position of the wave front defined as the point where $u$ goes
above 0 and $x_{off}$ is an arbitrary value lower than $L$ and
significantly bigger than the wave front width. The following
boundary conditions are applied:
\begin{equation}
\left\{
\begin{array}{l}
\mbox{at $x=x_{max}(t)$: }u=u_0,\, v=v_0\\ \mbox{at $x=x_{min}(t)$:
}\left\{
\begin{array}{lcl}
\partial_x u(0)=0 &\mbox{ if }& x_{min}(t)=0\\
\delta_u(0)=0      &  \mbox{ if }& x_{min}(t)\neq 0
\end{array}
 \right.
\end{array}
\right.
\end{equation}
The condition $\delta_u =0$ implies that the recovery past $x_{min}$
is purely deterministic and hence does not contribute to $E$. A check
on our procedure is afforded by the fact that as long as the distance
the wave had traveled (in the final state) is significantly bigger
than the region over which the wave initiates, the results obtained
are independent both of the distance the wave had traveled and of the
width of the space window used ($L$). We fix $T$ to be large enough
that the deviation from deterministic dynamics for very small time is
negligible. Once $T$ is fixed and the specific final wave state
chosen, there is no time-translation invariance in the MPEP.
\begin{figure}
\begin{center}
\hspace{1.cm} \includegraphics[width=4.0cm]{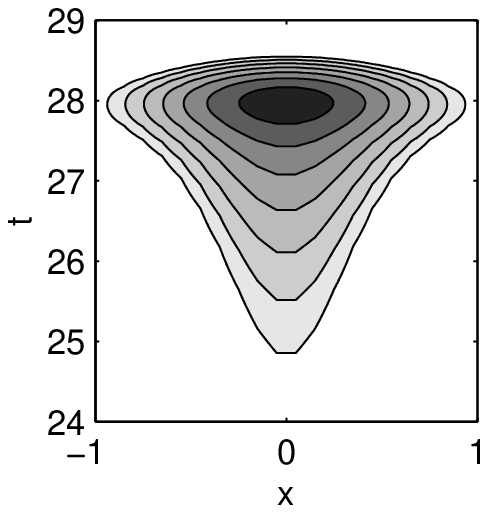} \newline
\includegraphics[width=3.5cm]{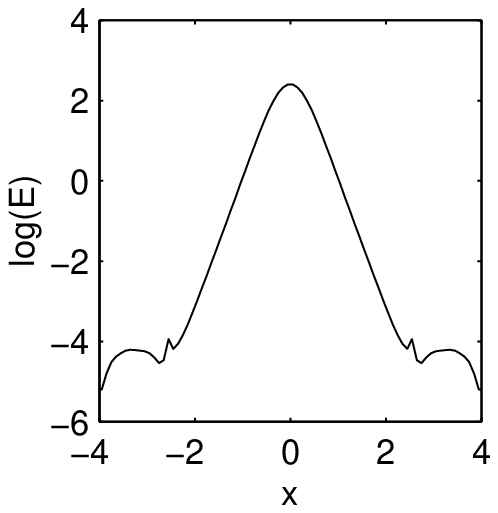}
\includegraphics[width=3.5 cm]{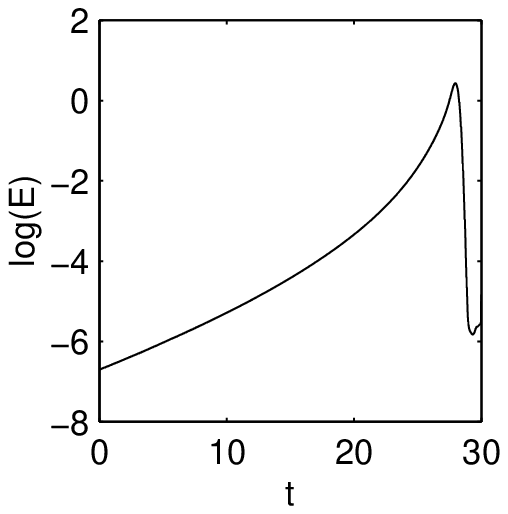}
\caption{\label{Erreur} Parameters values are $\epsilon=0.1$ and
$\gamma=.2$. Space time contour plot of the action density
$\delta_u^2+\delta_v^2$ with its maximal value normalized to one. The
contours are at  $ 1/96,\ 1/48,\  1/24,\ 1/12,\ 1/6,\ 1/3\mbox{ and }
1/1.5 $. (b)  Left: spatial distribution of the contribution to
$\log_{10}(E)$ during the MPEP; Right: temporal distribution of the
contribution to $\log_{10}(E)$ during the MPEP. One can see that the
contribution to $E$ as a function of time increases exponentially and
then decreases rapidly to a very low level that is close to numerical
noise }
\end{center}
\end{figure}

\begin{figure}
\begin{center}
\includegraphics[width=3.5cm]{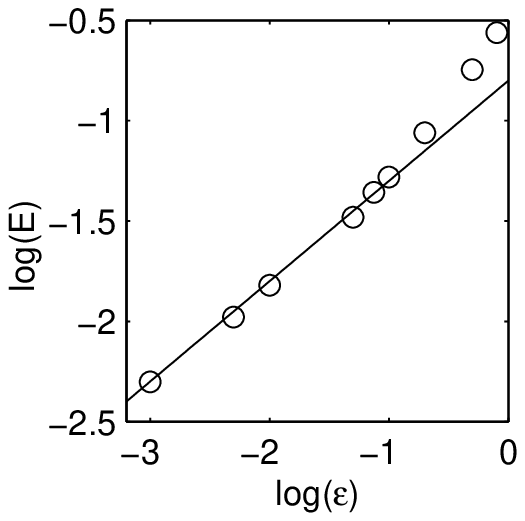}
\includegraphics[width=3.5cm]{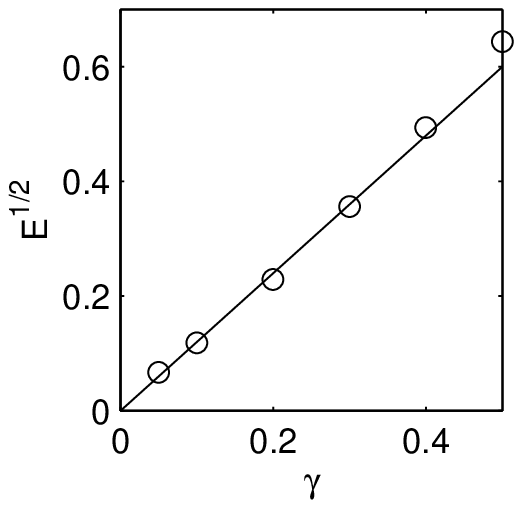}
\caption{\label{dependance}Left: circles $\log_{10}\epsilon$
dependence of $\log_{10} E$ using a log  scale, solid line:
$y=\frac{x}{2}-0.8$. For small $\epsilon$, $E$ behaves like
$\sqrt{\epsilon}$.  Right: circles $\gamma$ dependence of $\sqrt{E}$
for $\epsilon=0.1$. Solid, $y=1.2\, x$ $E$ behaves like $\gamma^2$
for a wide range of values of $\gamma$. The scaling breaks down for
higher values of $\gamma$, close to the propagation threshold  (for
this value of $\epsilon$, the threshold is at
$\gamma=0.5525\pm0.0005$).}
\end{center}
\end{figure}

We now describe the results obtained using this methodology.   In the
small $\epsilon$ limit ($\epsilon< 0.1$), the shape of the MPEP is
quantitatively independent of $\epsilon$ and $\gamma$ and even for
high values of $\epsilon$, there is no significant qualitative
difference. In fig. \ref{DessinMPEP1}, we present such a typical
escape path. As shown on the contour plots, wave nucleation is found
to be a very localized event. Essentially, the noise acts to create a
local dip in value of the $v$ field which is then followed by a large
positive excursion for the $u$ field as it goes into the excited
phase. To describe this mechanism more fully, we present in fig.
\ref{DessinMPEP2} the phase-plane trajectory at the center of
nucleation as well as several snapshots of the spatial form of the
 fields during the nucleation process. One can see then that the
escape path consists of the center of nucleation being driven by
noise below the minimum of the $f$ nullcline; this then quickly
drives the $u$ field positive and leads after, $v$-field driven
relaxation, to the pair of counter-propagating pulses.  There is a
significant fluctuation contribution to the nucleation event in a
small region around the nucleation point(see fig. \ref{Erreur}).
Results using different values of $\gamma$ and $\epsilon$ show that
the width of that small region is proportional to the width of a
front, that is $\sqrt{\epsilon}$ (see fig. \ref{dependance}). Note
that the other simple possibility, that of nucleating an excited
region for the $u$ field at a fixed value of $v$\cite{Pumir},
is not observed.

In accord with the shape of the trajectory, our calculations show
that the main contribution to $E$ during the MPEP come from the
$\int\int \delta_v^2$, at least for small values of $\epsilon$. Thus
for $\gamma=0.2$ the value of the ratio $\int\int \delta_v^2/\int\int
\delta_u^2$ never went below 30 for $\epsilon=0.1$, and went up to
1000 for the minimal value of $\epsilon$ used ($\epsilon=0.001$). For
higher values of $\epsilon$, the ratio was significantly lower (3 for
$\epsilon=0.8$, close to the limit of propagation for this value of
$\gamma$ ). Furthermore for small $\epsilon$,  $E$ scales like
$\sqrt{\epsilon}$ (see fig. \ref{dependance}). For higher values of
$\epsilon$, this simple scaling is no longer valid. The fact that for
small values of $\epsilon$, $E$ scales like $\sqrt{\epsilon}$ and
that a wave is therefore much easier to nucleate can be explained by
a simple argument.
As already mentioned, the phase-plane trajectory of the MPEP is mainly
independent of the value of $\epsilon$. The only dependence is then
due to the spatial scale appearing in the integral which is
$\sqrt{\epsilon}$, giving rise to the observed result

We now describe results obtained when varying $\gamma$ with
$\epsilon$ held constant. This means that we consider the nucleation
rates for differing excitabilities but with the same time-scale ratio
between the $u$ and $v$ field dynamics. For a wide range of values of
$\gamma$, we find that $E$ scaled like $\gamma^2$ (see  fig.
\ref{dependance}). The more excitable the system, the more likely it
is to nucleate a wave. This scaling is that same as one would obtain
analytically in the much simpler  zero-dimensional version of
this problem. Here the analog of wave nucleation is the noise-induced
creation of a large transient excursion away from the fixed point. If
$\epsilon$ is small, the escape path will follow the $f$ nullcline
down to its minimum and then follow the noiseless dynamics to reach
the other stable branch. In such a situation, one can compute the
transition rate analytically in a piece-wise linear version of the FH
model,
\begin{eqnarray}
\dot{u}&=&\left\{
\begin{array}{ll} -\frac{1}{\epsilon} (v+u) &\mbox{if $u <0$},\\
\frac{1}{\epsilon}( 1-v-u) & \mbox{if $u\ge 0$},
\end{array}
\right.\\
\dot{v}&=&u-\gamma.
\end{eqnarray}
After elimination of $u$, it can easily be shown that the variational
equation for the MPEP has the form
\begin{equation}
\frac{d^2 v}{dt^2}=v+\gamma
\end{equation}
with $v(0)=v_0=-\gamma$ and $v(T)=0$. A straightforward calculation
shows that the action is equal to
\begin{equation}
\gamma^2\frac{1}{2(1-\exp(-2T))^2}
\end{equation}
and that its minimum, reached for $T=\infty$, is equal to
$\gamma^2/2$. This result shows that the $\gamma^2$ scaling of $E$
found in our computations can be interpreted as being due to $\gamma$
equaling the ``distance'' of the stable equilibrium point from the
border of its basin of attraction. Putting it all together, our data
yield a log transition rate proportional to $-\frac{\gamma^2
\sqrt{\epsilon}}{\beta}$. This result could be tested experimentally,
perhaps by adding illumination noise to the light-sensitive BZ
reaction.

It is worth mentioning that there are other potential applications of
the MPEP approach to nucleation in spatially-extended non-equilibrium
systems. One example concerns the thermal generations of localized
patches of traveling rolls in electro-convection, as studied
recently\cite{Bisang}. Also, the method used here is not limited to
white noise. A simple generalization of the derivation allows for the
incorporation of multiplicative noise via dividing the $\delta_u^2$
and $\delta_v ^2$ terms in the action density by the corresponding
(possibly field dependent) variances of the noises added to the $u$
and $v$ equations respectively. This will clearly be necessary for
the study of realistic calcium models. Finally, there is a similarity
between the MPEP method and what must be done to consider quantum
tunneling in spatially extended systems\cite{Freire}, where one also
must find the entire space-time path tunneling trajectory in order to
find a leading estimate of the rate.

It is a pleasure to acknowledge useful discussions with D. Kessler.
This research is supported by the National Science Foundation through
Grant No. DMR-0101793.

\bibliography{herve}

\end{document}